\documentclass[runningheads]{llncs}
\usepackage{graphicx}
\usepackage{booktabs}
\usepackage{float}

\begin{document}
\title{Patient Perspectives on Telemonitoring during Colorectal Cancer Surgery Prehabilitation\thanks{Supported by ITEA 3 Inno4Health 19008 Project}}

\titlerunning{Patient Perspectives on Prehabilitation Telemonitoring}
%
\author{Irina Bianca Șerban\thanks{Corresponding author}\inst{1} \and
Dimitra Dritsa \inst{1} \and
David ten Cate \inst{2} \and
Loes Janssen \inst{2} \and
Margot Heijmans \inst{2} \and
Sara Colombo\inst{3} \and 
Aarnout Brombacher\inst{1} \and
Steven Houben\inst{1}} 

\institute{Institute}
\institute{Eindhoven University of Technology, Eindhoven, Netherlands \\ \email{i.b.serban@tue.nl} \and
Máxima Medical Centre, Veldhoven, The Netherlands \and
Delft University of Technology, Delft, Netherlands
}
\authorrunning{Șerban et al.}
%
%
\maketitle              

\begin{abstract}
Multimodal prehabilitation for colorectal cancer (CRC) surgery aims to optimise patient fitness and reduce postoperative complications. While telemonitoring’s clinical value in supporting decision-making is recognised, patient perspectives on its use in prehabilitation remain underexplored, particularly compared to its related clinical context, rehabilitation. To address this gap, we conducted interviews with five patients who completed a four-week CRC prehabilitation programme incorporating continuous telemonitoring. Our findings reveal patients’ willingness to engage with telemonitoring, shaped by their motivations, perceived benefits, and concerns. We outline design considerations for patient-centred systems and offer a foundation for further research on telemonitoring in CRC prehabilitation.

\end{abstract}
\section{Introduction}

Colorectal cancer (CRC) is the second leading cause of cancer-related deaths worldwide~\cite{Torre2015CRC}. Surgery is the main treatment for non-metastatic CRC, but postoperative complications often prolong hospital stays, delay recovery, and increase costs~\cite{Rooijen2019PrehabForCRCRCT,McSorley2016PostoperativeComplicationsCRCSurgery}. Rehabilitation can mitigate these complications~\cite{Nusca2021RehabSurgeryCRC}, yet it cannot fully reverse the significant physiological and functional decline -- up to 40\% -- that many patients experience~\cite{Cuadros2017HRMonitoringPrehab,Jakobsson2014RecoveryAfterCRCSurgery}. This has shifted attention toward \textit{prehabilitation}: proactive, multimodal interventions to boost functional capacity and resilience \textit{before surgery}~\cite{Silver2015CancerPrehab}.

Emerging as a distinct phase in the cancer care continuum ~\cite{Steffens2023PerspectivesOnlinePrehab,Molenaar2023Prehab,vanRooijen2019MultimodalPrehabPositionPaper}, prehabilitation combines in-clinic sessions with at-home health behaviour guidance. Unlike rehabilitation, it demands intensive engagement over a compressed timeframe -- typically one-third the duration~\cite{Falz2022EffectsDurationPrehab} -- and aims to optimise surgical outcomes rather than promote long-term recovery~\cite{Silver2015CancerPrehab}. For CRC patients, prehabilitation begins immediately after diagnosis, amid emotional upheaval, intensive programme requirements, and life disruption~\cite{Beck2021PatientPerspectivesPrehabilitation}. These unique pressures pose distinct technological challenges~\cite{Zhu2020TechnologicalOportunitiesPrehab}.

Although rehabilitation is well studied in Human-Computer Interaction (HCI) \cite{Kjaerup2018DiagnosticAGentsTelerehab,Shimbo2023TelemonitCardiacRehab,Kabir2019RemoteMonitoringCancer,Piotrowicz2012TelemonitoringRehab}, little research explores prehabilitation’s technological needs from the patient perspective~\cite{Zhu2018TechnologyInPrehab}. Continuous telemonitoring of patient behaviour and well-being has been proposed to support clinicians in managing diverse patient needs~\cite{Franssen2022TeleprehabFeasibility,Cos2022TelemonitoringPrehab,Cuadros2017HRMonitoringPrehab,Rumer2016WearablesPrehab}, with studies prioritising technical or clinical outcomes (e.g.,~\cite{Cos2022TelemonitoringPrehab,Cuadros2017HRMonitoringPrehab}), partly due to the challenges of involving patients during this intense phase (e.g.,~\cite{Franssen2022TelePrehab}). Yet, understanding patient perspectives is critical: telemonitoring must be non-intrusive and offer clear value to foster adherence and improve surgical outcomes~\cite{Fan2024PatientPerspectivesRCTTMCancer}. Although pervasive technologies can be ubiquitous in the patient’s home environment~\cite{Cuadros2017HRMonitoringPrehab,Rumer2016WearablesPrehab}, poor design risks disrupting their routines, particularly in the high-pressure context of prehabilitation. Systems must engage patients meaningfully without adding burden -- a complex challenge, as what constitutes \textit{meaningful engagement} in this setting is unclear. 

To address this gap, we explore (i) \textit{CRC prehabilitation patients’ experiences with telemonitoring} and (ii) \textit{their perceived values and potential uses for telemonitoring in managing health during prehabilitation}.  Through qualitative interviews with five participants completing a 4-week programme involving on-body sensors and self-reporting tools, we contribute: (i) an empirical overview of patients’ experiences with telemonitoring during CRC prehabilitation, (ii) insights into perceived added value and motivators, and (iii) a discussion to inform the design and research of future patient-centred telemonitoring systems for prehabilitation.
\section{Related Work and Motivation}

Prehabilitation, the first phase of the cancer rehabilitation continuum, occurs between diagnosis and surgery to improve health and reduce post-surgical complications~\cite{Silver2015CancerPrehab}. Even without complications, major surgery often leads to significant functional decline~\cite{Jakobsson2014RecoveryAfterCRCSurgery}. Prehabilitation is also used in abdominal~\cite{barberan2018personalised}, lung~\cite{Cate2024PrehabLung}, and pancreatic cancer surgeries~\cite{van2020pancreatic}. Despite advances in care, CRC surgery still shows high 30-day morbidity rates (19\%–37.4\%)~\cite{Sharp2020OutcomesRectalCancer},  influenced by factors like functional capacity, mental health, nutrition, and smoking. \textit{Multimodal prehabilitation}  programmes combine physical training, psychological support, dietary guidance, and smoking cessation (e.g.,\cite{Molenaar2023Prehab,Rooijen2019PrehabForCRCRCT,Basquet2018MultimodalPrehab}), typically over 4–6 weeks~\cite{LEROY2016Prehabilitation}, delivered by multidisciplinary teams (nurses, physiotherapists, dietitians)~\cite{Renouf2022InterdisciplinaryPrehab}. 

Personalised prehabilitation is essential for inrcreased adherence and stress reduction~\cite{Beck2021PatientPerspectivesPrehabilitation}, but clinicians face challenges tailoring interventions within compressed timelines~\cite{Zhu2020TechnologicalOportunitiesPrehab,zhu2018ChronicPrehab}. Heavy workloads mean clinicians often rely on visit-based data, limiting insights into at-home behaviours~\cite{Zhu2020TechnologicalOportunitiesPrehab}. ICT systems like on-body sensors~\cite{Rumer2016WearablesPrehab,Cos2022TelemonitoringPrehab}, and self-reporting apps~\cite{Sliwinski2023ToolboxPrehabSurgery} have been proposed to bridge gaps in patient-centred cancer care~\cite{Clauser2011ICTCancer}, supporting both telehealth ~\cite{Fan2024PatientPerspectivesRCTTMCancer,Doiron2020TelePrehab,Franssen2022TelePrehab}, and in-clinic settings~\cite{Cos2022TelemonitoringPrehab,Sathe2018GaitAnalysis}.

While clinicians’ technological needs are increasingly studied (e.g.,\cite{SerbanDritsa2023DataNeedsPrehab,Zhu2018TechnologyInPrehab}), patient perspectives -- especially in HCI -- remain underexplored~\cite{Zhu2020TechnologicalOportunitiesPrehab}. As CRC prehabilitation is relatively new~\cite{Rooijen2019PrehabForCRCRCT}, research into patient-centred telemonitoring remains sparse. The value of telemonitoring for clinicians in prehabilitation is established, supporting assessment and decision-making~\cite{SerbanDritsa2023DataNeedsPrehab,Zhu2020TechnologicalOportunitiesPrehab}. However,  it is patients who interact with these systems daily. For successful engagement, telemonitoring must be comfortable, easy to use, and deliver clear value~\cite{Andersen2017PatientExperienceTM}, aligning with User Experience (UX) research in HCI, which emphasises subjective, emotional, and contextual dimensions of technology use~\cite{Bargas-Avila2011UserExperience,HassenzahlTract2006UX}. UX research in prehabilitation telemonitoring remains limited. Most studies focus on feasibility or clinical outcomes~\cite{Sathe2018GaitAnalysis,Doiron2020TelePrehab,Franssen2022TelePrehab,Cos2022TelemonitoringPrehab,Steffens2023PerspectivesOnlinePrehab,Doiron2020TelePrehab,Cos2022TelemonitoringPrehab,Rumer2016WearablesPrehab}, with few exceptions like Fan et al. ~\cite{Fan2024PatientPerspectivesRCTTMCancer}, who highlight proactive patient mindsets fostered by telemonitoring, and Cuadros et al.~\cite{Cuadros2017HRMonitoringPrehab}, who stress usability for older adults. Deeper investigations into patient needs are urgently required.

By contrast, related clinical contexts like cancer rehabilitation telemonitoring engages more with patient perspectives. Cerna et al.~\cite{Cerna2020DecisionSupportCancerRehab} identified challenges with self-report reliability, and Rossen et al.~\cite{Rossen2020PatientPerceptionCancerRehabTech} demonstrated how self-tracking supports physical activity. However, these insights come from rehabilitation, which focuses on gradual behaviour change~\cite{Nusca2021RehabSurgeryCRC}, whereas prehabilitation is fast-paced, emotionally charged, task-heavy and anticipative of a life-saving intervention~\cite{Fan2024PatientPerspectivesRCTTMCancer,Idenburg2020PAinolderAdultsInPrehab}. Despite the demands of prehabilitation, little is known about what motivates patients to engage with telemonitoring, how they perceive its value, or how best to design these systems. Does telemonitoring add to their burden? Should patients access their data, or should it remain clinician-facing? What devices and interactions suit this critical period? Our study provides early insights into these questions and underscores the need for further patient-centred research in cancer prehabilitation.


\begin{figure}
\includegraphics[width=\linewidth]{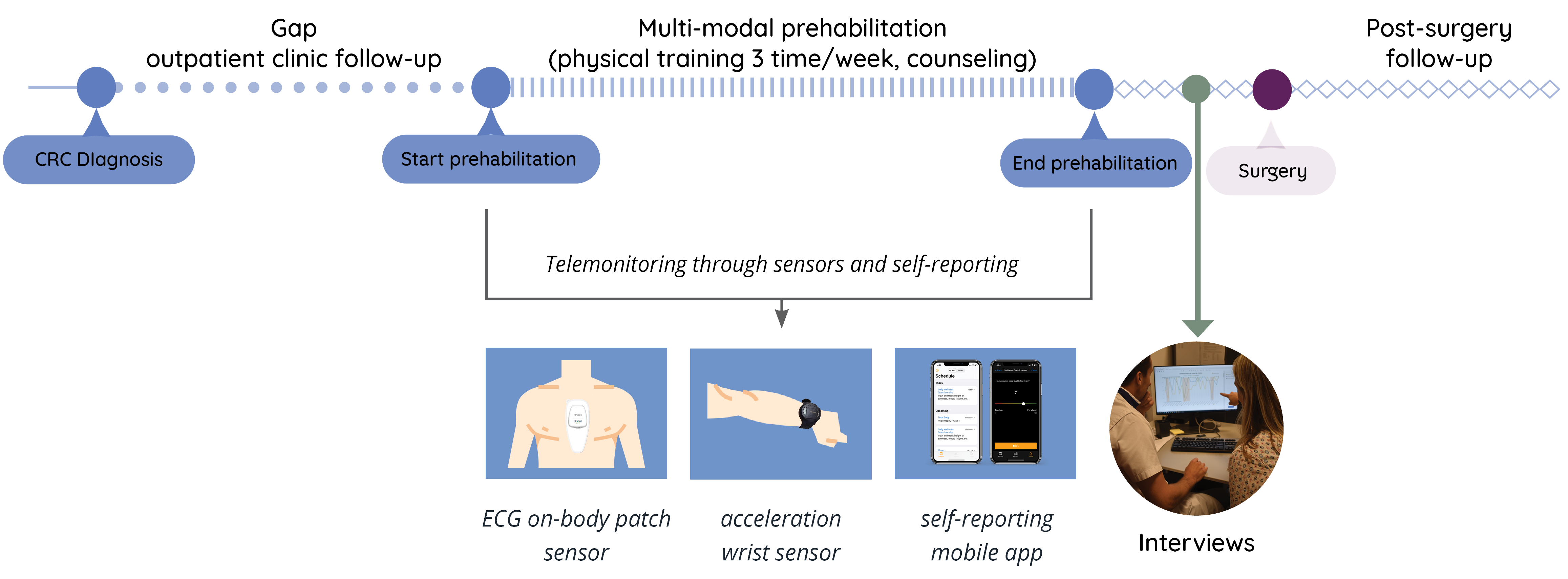}
\centering
\caption{Prehabilitation trajectory and set-up of the study.}
\label{fig:prehab}
\end{figure} 

\section{Study Context and Methods}

This study explores the experiences of five patients (P1-P5, median age 67, one female -- further details provided in Appendix~\ref{participants}) who underwent 24/7 telemonitoring during a CRC prehabilitation programme at a teaching hospital in the Netherlands -- Máxima Medisch Centrum Veldhoven. Participants wore an ECG sensor (Philips BioTelemetry e-Patch~\cite{biotel}), a wrist accelerometer (GENEActiv sensor~\cite{geneactiv}), and used a mobile app (Kinduct Lifestyle App~\cite{kinduct}) for daily logging of sleep, stress, motivation, barriers and activity at home via a 5-minute survey; a paper survey was available for backup. ECG sensors (10-day battery) were replaced bi-weekly, while wrist devices (four-week battery) required no replacement. System interaction was limited to app use and marking training sessions via the wrist device and the on-body sensors through button clicks. This setup was designed for a broader study, aiming to assess the feasibility of collecting continuous telemonitoring data to inform clinicians about functional capacity, activity, and well-being. Ethical approval was not required under Dutch law (METC N22.066), but the study was approved by the Institutional Review Board of Máxima
Medical Center (2023-MMC-015).

This feasibility study context enabled exploration of patients’ telemonitoring experiences alongside in-clinic prehabilitation and perceptions of its value. Each participant completed a four-week multimodal programme, starting 1–2 weeks post-diagnosis, with surgery scheduled immediately after. Patients were informed by a specialist nurse and consented to participate. The programme included three supervised physiotherapy sessions weekly, plus educational, dietary, and psychological support if needed. Following final training, participants completed a 45–60 minute open-ended exit interview (Figure~\ref{fig:prehab}). \textit{This paper focuses on patients’ telemonitoring experiences extracted from the interviews}. Of 71 screened patients, 21 were ineligible (for medical reasons, scheduling, or language barriers); of 50 eligible patients, 19 declined (mainly due to perceived burden and increased stress), 24 trained outside the hospital, and 7 enrolled; 5 participated in recorded interviews.

Semi-structured interviews explored: (1) experiences with telemonitoring and prehabilitation, (2) motivations and perceived value, and (3) potential applications for health management (Appendix~\ref{interview}). None had prior wearable experience, although P2–P4 occasionally used smartphone step counters. The system only gathered data, and participants had no real-time data access during the study. To support reflection, interviews included visual aids (Appendix~\ref{visual}): (1) data cards on stress, fatigue, sleep, and activity, and  (2) sample visualisations based on each patient's data, linking self-reported physical activity, stress, motivation and barriers. This approach facilitated discussion around the perceived value of telemonitoring during prehabilitation. All interviews were conducted face-to-face, audio-recorded, and transcribed verbatim. Data analysis followed a hybrid deductive and inductive thematic analysis~\cite{Braun2006ThematicAnalysis}, supported by NVivo 14 (QSR International, Melbourne, Australia).

\section{Findings}
We first outline \textit{general perceptions of the prehabilitation program} to provide context, then describe \textit{experiences with telemonitoring}, including attitudes and social perceptions. We highlight \textit{perceived values of telemonitoring}, its envisioned \textit{applications for supporting health management}, and describe \textit{trade-offs} such as mistrust in the healthcare system.

\subsection{Perceptions on CRC Surgery Prehabilitation}

Participants described the profound impact of diagnosis, highlighting emotional strain, feelings of resignation, and the need to put daily life and future plans on hold.  P1, for example, spoke of postponing travel and adjusting energy levels to manage prehabilitation. Balancing prehabilitation with preparations for life post-surgery was also a challenge; P4 aimed to complete roof repairs before surgery despite fatigue from physiotherapy: \textit{``I have in my mind that I want to finish that roof, before I get my operations [...] But then I am still tired from the fysio, and it takes a bit longer...''}

Despite difficulties, participants saw surgery and prehabilitation as a glimmer of hope, offering a sense of control amidst uncertainty. P3 reflected: \textit{``I can start crying, but that does not really help, does it? [...] My reasoning is, they discovered it, probably in time, it is isolated, it is operable, and they fixed me.''}

All participants expressed satisfaction with the received care, valuing group training and the sense of community. P5 remarked: \textit{``If you come here, you are never a number. You are always a human being.''} Though four participants acknowledged the program's intensity, which could cause fatigue alongside life disruptions, they described feeling proud and motivated by their observable physical progress. For instance, P5 shared: \textit{``The day before yesterday for example we walked for more than 1.5 hours and then I was not tired at all, while I used to be tired before.''} The program’s intensity built trust in their bodies and encouraged perseverance. Participants felt motivated and secure due to frequent clinician contact and supervision, which provided reassurance and stability during a period of significant uncertainty. P2 described this as a source of comfort: \textit{``I think it's also a luxury you've got. I always said this when this started: I am surrounded by care. It's incredible.''} The program's structured nature, with its dense schedule, and clear goal of surgery, further motivated participants to remain active at home through walking, exercising, and household activities.

\subsection{Experiences with Continuous Telemonitoring}

\subsubsection{General Attitudes -- Comfort, Usability and Willingness to Share Data.}
Overall, participants did not report negative experiences with the telemonitoring system. Attitudes were positive or neutral, with the system seen as non-intrusive both physically (e.g., not restricting movement, sleep or causing discomfort) and in daily life (e.g., not requiring disrupting interactions -- e.g., charging, logging). Patients reported not feeling supervised or uncomfortable, despite being monitored 24/7. Patch comfort and discreetness contributed to this, with P5 even forgetting about wearing it. P2 and P3 experienced mild skin irritation, but found it unobstrusive: \textit{``It did not matter. The first few days, I had a little irritation of the skin; itching, but that is it''} (P3).

Completing the self-reporting questionnaires was considered manageable and not overly burdensome. P4 used the paper survey when they forgot to use the app, explaining that they do not consistently engage with their mobile phone throughout the day. Participants had no privacy concerns or reluctance to share data, provided that data access was limited to their healthcare institution. They expressed feeling comfortable sharing the telemonitored data at any time with their healthcare providers during prehabilitation, motivated by trust in the institution, a desire to do everything possible for their health, and a willingness to support healthcare professionals.

\subsubsection{Social Perceptions of the Telemonitoring System.}
Four participants noted that friends or family did not comment on the telemonitoring system. They suggested that the system was either discreet and therefore unnoticed, or, when noticed, seen as an integral part of prehabilitation and met with encouragement or neutrality. P4 highlighted that their social environment viewed the sensors and the app as a natural aspect of preparing for surgery, further reinforcing the lack of negative attention or commentary from others:\textit{``I think that no one really cares [...] They just know an operation is about to happen [...] This operation can have some really good effects or results, that I might benefit from later. And this [telemonitoring] can only help to get even better.''} This social acceptance contributed to participants' overall positive experience with the system.

\subsection{Values and Motivations for Engaging in Telemonitoring}

\subsubsection{Gaining Agency Over Diagnosis.}
Participants described a profound loss of control following their cancer diagnosis, including a perceived lack of agency over surgical outcomes. Engaging in prehabilitation, however, restored a sense of empowerment, allowing them to take proactive steps towards improving their health. Participants expressed a strong willingness to do anything to optimise their surgical condition, which shaped their positive perceptions of telemonitoring. P5 framed participation as essential: \textit{``You sign up and you just participate, because it is for yourself. I don't think as much about it, it just has to happen, you just do it.''} Telemonitoring was seen as an integral part of prehabilitation and valued for its role in supporting health optimisation. P2 similarly remarked: \textit{``If it is necessary I would do anything. Does not matter.''} P4 noted their partner’s support for telemonitoring, recognising its potential to improve surgical outcomes: \textit{``She just thinks that it is good for me to enter the operation in an optimal state and the more support to achieve that, the better.''} Telemonitoring was thus willingly accepted as a tool for enhancing physical readiness for surgery.

\subsubsection{Supporting Clinical Care Through Telemonitoring.}
Participants also expressed a desire to contribute value to the clinicians offering them life-saving supporting. By providing telemonitoring data, they felt they could assist healthcare professionals in delivering better care, streamlining their work, and improving outcomes. P5 articulated this sense of gratitude, viewing participation as a way to give back: \textit{``They found my cancer with the population screening and that happened because they are doing a lot of research. And then I am asked to do something back and I think, yeah then I also do something in return, because if that [population screening] never happened then I would maybe not be there anymore in two years.''} This sense of thankfulness motivated her to give back to the healthcare institution by contributing through telemonitoring, knowing it could benefit the practitioners supporting her recovery. Similarly, P4 saw value in sharing data with clinicians if it helped them in their work, ultimately benefiting the patient as well: \textit{``it actually does not matter to me. I don't have any secrets, so if maybe it will help the doctor, then sure, just do it''}. Their willingness to engage reflected trust in the healthcare system and a belief in the mutual benefits of collaboration.

\subsection{Envisioned Applications for Telemonitoring}

\subsubsection{Receiving Data-Driven, Personalised Support. }
Patients recognised that sharing telemonitoring data with clinicians could enhance care by supporting decision-making and improving treatment. Participants agreed that such data could offer better insight into behaviour outside the clinic, leading to more tailored advice. As P3 noted: \textit{`` I don't mind that the doctor can see it. [...] I assume that it can even contribute to a better diagnosis. Once I have a complaint or an issue they might be able to make a better adjustment.''} Similarly, P1 expressed openness to sharing activity data from home with their physiotherapist and surgeon, hoping it could contribute to improved treatment outcomes.

Participants envisioned two main applications for telemonitoring: patient-initiated help and clinician-initiated action. In the first, patients could use data like high stress levels to seek support proactively. P2 illustrated this: \textit{``When my stress level would be very high, I would call [nurse's name] [...] And then she says `There is this relaxation program on Youtube'. ''} In the second, participants valued clinicians monitoring trends and reaching out when sustained anomalies appeared, emphasising the need for action based on persistent patterns rather than isolated events. As P5 explained: \textit{``I don't think that after one peak action should be taken immediately. In the case that I would go down [be very fatigued] and stay down for example, then action is needed. But I, and everyone can have a peak sometimes, so not immediately.''} P2 agrees: \textit{``When it would be three days, or four, then it would be a really nice thing for them to call you because something is happening.''}

\subsubsection{Improving Self-Reflection and Decision-Making.}
Participants were open to interacting with telemonitoring data if it added value during prehabilitation. They saw potential in sensor and self-reported data to enhance self-awareness, reflect on their actions, validate internal feelings, and identify behavioural improvements. They valued data as a tool to confront themselves and balance subjective perception with objective facts. For instance, P2 found that logging daily activities like chores increased awareness of their physical effort beyond formal exercise. They also felt stress data could help them confront avoidant coping habits and \textit{``look into the mirror''}. P3 similarly noted that daily self-reporting fostered greater awareness of inactivity: \textit{``Sometimes you sit down and you are being lazy [...] You have to do this [self-report activities] on a daily basis, so you face it. You are getting a little bit more awareness.''} 

Some participants were surprised when shown their telemonitoring data, which challenged assumptions about their behaviour and offered encouragement. P5, who initially felt they were underperforming, realised they were consistently active with chores -- activities they usually overlooked: \textit{``I am relieved to hear that.. I always think of myself as lazy.''} Participants agreed that seeing their data could validate feelings and recalibrate their self-perception: \textit{``Use it for managing my health, understanding my body, then I would want that''} (P3). P5 suggested weekly or bi-weekly reviews to better assess their progress: \textit{``Am I really lazy, or do I just think that? I would like to know that.''} Similarly, P3 noted that accessing stress data could enhance emotional grounding during a period of heightened activity and uncertainty: \textit{``Maybe it is not about doing things differently but being conscious about what is happening. I think that would help and motivate people.''} 

Participants also expressed a desire to act on insights from their data. Balancing the wish to stay active with managing fatigue was a common challenge. P1 highlighted the benefit of seeing correlations between activity, sleep, and fatigue to know when to rest: \textit{``I am feeling myself pretty well, but then I see I am fatigued and I understand I have to slow down a bit.''} Similarly, P4 valued understanding fatigue triggers: \textit{``What are the circumstances in which I am tired or more easily tired and what can I do about it?''}. P5 noted that activity data could guide adjustments in individual or joint activities with their partner to boost activity levels when needed.

\subsubsection{Enabling Informal Caregiver Participation}

Participants valued involving informal caregivers, like partners, as active collaborators in telemonitoring, believing it could strengthen communication and their shared health journey. They saw opportunities for caregivers to assist with data logging and daily reflection. For instance, P1 described completing questionnaires together in the evening, sparking discussions about the responses: \textit{``Most of the time at night at 8 o'clock she is there, because we sit at the TV. [...] We discuss from time to time about the questions like, ‘What do you think about this?''} Similarly, P4 envisioned sharing objective data such as sleep patterns with their partner to bridge the gap between their subjective feelings and their partner’s perception: \textit{``My feeling and her feeling are never truly aligned right?''}

Participants also saw telemonitoring as a way to ease caregiver anxiety through objective reassurance. P3 noted that sharing activity data could address their partner’s concerns about inactivity. When faced with their self-reported activities, P3 says: \textit{``Tell my wife! She thinks I can be more active. [...] She is more concerned about the fact that I may sit down and wait for my operation.''} Meanwhile, P4 described their partner as overprotective, sometimes leading to tensions when P4 wished to push harder. In such cases, fatigue data could help mediate conflicts: \textit{``She protects me. Well, these are strange times, she is more intense than usual.''} P2 added that stress data could help negotiate responsibilities, describing how they communicated the need to hand over household chores when energy levels were low: \textit{``I told my partner, from now on, I stopped doing the household, you must take it. Cause I think it’s bad. And not because I am lazy or something. I need to take my rest.''}

Finally, participants stressed the importance of maintaining autonomy in sharing data with caregivers. P5, for instance, valued their partner’s concern but preferred to control when and how information was shared, opting to consult their doctor first if issues arose: \textit{``I am responsible for my own health.''}

\subsection{Trade-offs and Barriers of Telemonitoring}

\subsubsection{The Dangers of Added Burden and Limited Utility.}

Despite the envisioned benefits of telemonitoring for health management, participants highlighted trade-offs that could limit its utility and add burden during an already intensive programme. The primary interaction was through the self-reporting mobile app. P1 found the daily surveys occasionally excessive, particularly given the programme’s intensity and frequent hospital visits. They noted the surveys did not fully accommodate daily life variations (e.g., social interactions or fluctuating energy levels), and limited multiple-choice options often forced them to use open fields, adding extra work. Similarly, P2 found open-field reporting tedious and preferred structured multiple-choice responses, seeing little self-management value from detailed activity labelling.  

While participants acknowledged telemonitoring’s potential, some questioned its practical value. P1 and P3 felt they already had a strong internal compass and were sceptical whether, in routine practice, additional data would improve their health management. P2 preferred relying on memory during consultations, noting that fitness reports already available through physiotherapy lessened the need for further data. P4 and P5 stressed that any technology must provide clear, demonstrable value to both themselves or their clinicians.

Some participants were also concerned about the medicalisation of daily life. P3, for instance, valued keeping home life separate from hospital routines and viewed telemonitoring as potentially intrusive, especially for private activities like enjoying a weekend beer, which conflicts with prehabilitation guidelines. P2 further raised concerns about feelings of shame when low activity levels might be revealed to clinicians.

\subsubsection{Scepticism About Clinicians Acting on Telemonitoring Data. }

Some participants expressed concerns about whether clinicians would appropriately act on telemonitoring data, rooted in past negative experiences of being ignored or misdiagnosed. These experiences shaped a sense of scepticism and mistrust in clinicians' responsiveness to patient-reported issues. P3, for example, questioned whether clinicians would respond to alarming signals, citing frustration over a previous misdiagnosis of bowel cancer and a belief that cost-saving measures had compromised care: \textit{``I sometimes am a little critical about doctors. Because I mentioned some problems earlier to my doctor, and they said `It is irritable bowel syndrome.' [...] That makes me believe that doctors are trying to save costs in the wrong way.''} This mistrust extended to telemonitoring, as they believed similar dismissive behaviour could occur remotely.

Similarly, P4 recounted a long history of being ignored by their general practitioner for a different health issue and questioned whether clinicians would genuinely take action if the telemonitoring data flagged a concern. They also raised a related concern about the potential for data overload, suggesting that excessive information might dilute the clinician's sense of urgency or responsibility to act: \textit{``You should not overload him with information, because then the feeling of 'I have to do something about this' becomes less.''}

\section{Discussion}

Our study highlights the need for telemonitoring systems in CRC prehabilitation to be both clinically effective and patient-friendly. While telemonitoring supports clinician decision-making~\cite{Zhu2020TechnologicalOportunitiesPrehab,SerbanDritsa2023DataNeedsPrehab}, it must align with patient needs to foster trust and adherence. Participants stressed that telemonitoring should deliver meaningful value to their care, reinforcing calls for patient involvement early in system design and after system use~\cite{Serban2025CTRAthletes,Ramachandran2022Review}.

While wearable trackers have been proposed to boost prehabilitation activity~\cite{Rumer2016WearablesPrehab}, we find a more nuanced dynamic: our patients were already highly motivated by the goal of optimising surgical outcomes. Telemonitoring has the potential to reinforce patients' \textit{sense of control}~\cite{Powell2023PrehabonEmotionalWellbeing} rather than directly influence behaviour. Another envisioned valuable application for telemonitoring was its potential to support \textit{a collaborative model of care}~\cite{Fan2024PatientPerspectivesRCTTMCancer} -- the patients' desire to support clinicians in providing the best care, as well as being able to ask for and receive personalised advice. Furthermore, we observed that balancing program engagement with fatigue was challenging for our participants. Telemonitoring could assist by visualising fatigue and activity patterns, helping patients \textit{balance effort and recovery}. Beyond formal exercise, patients overlooked everyday activities like chores, which telemonitoring could highlight to \textit{boost morale and reinforce progress} against the emotional weight of the diagnosis~\cite{Idenburg2020PAinolderAdultsInPrehab}.

 Recent HCI research highlights the need for uncovering telemonitoring motivators between prehabilitation and rehabilitation~\cite{Zhu2020TechnologicalOportunitiesPrehab}. Our findings suggest that, unlike rehabilitation patients who use technology to sustain exercise behaviours over time~\cite{Rossen2020PatientPerceptionCancerRehabTech}, prehabilitation patients might engage with telemonitoring for a desire to maximise surgical success, maintain hope, feel agency and prevent physical stagnation. Given these strong short-term motivators and positive experiences, technology adherence in prehabilitation may exceed that in rehabilitation, where post-discharge engagement is hindered by the desire to return to normal life and the challenge of sustained behaviour change~\cite{Tadas2020Barriers}. However, further research is needed to confirm this and clarify what drives engagement in prehabilitation telemonitoring.

We also extend research on the emerging role of informal caregivers within telemonitoring systems (e.g.,~\cite{Serban2025CTR,Bhat2023InformalCaregivers}). Our patients valued involving partners in logging and interpreting telemonitoring data, seeing it as a way to mediate conflicts, improve support, and foster shared understanding. Future systems should facilitate caregiver collaboration and data participation.

System usability was another key theme. Participants appreciated non- intrusiveness, minimal maintenance, and short usage periods, echoing prior findings~\cite{Cuadros2017HRMonitoringPrehab,Huygens2021TelemonitUptake}. To ensure accessibility, systems must accommodate users less familiar with mobile technologies through alternative means of logging data. 

However, important trade-offs emerged. Frequent data collection, while clinically valuable, can overwhelm patients already managing intense schedules. Systems must align reporting frequency with patient capacity~\cite{Cerna2020DecisionSupportCancerRehab}, following principles like Andersen’s alignment framework~\cite{Andersen2018Alignment}. Mistrust in the healthcare system also surfaced. Patients doubted whether clinicians would act on telemonitoring data, citing past dismissals. Systems must foster trust through clear communication and responsiveness~\cite{Andersen2017PatientExperienceTM}. Additionally, some feared being judged for inactivity, supporting findings from Gupta et al. ~\cite{Gupta2020IdidntDoAGoodJob}. Systems should avoid punitive feedback and instead emphasise achievable goals and encouragement.

Lastly, participants voiced concerns about telemonitoring medicalising their home lives, seeking to protect spaces of normalcy. Minimising intrusion and offering opt-out options may be necessary to support emotional well-being, balancing clinical benefit with respect for the lived experience of cancer prehabilitation.

\subsubsection{Limitations and Future Work.}
Our participants did not access data or receive clinician feedback through the system, so future research should explore technologies enabling active patient-clinician-data interaction. Our system was designed for clinical data collection, and we leveraged this context, where recruitment is challenging, to capture patient experiences. Consequently, our methods of data collection did not always match the patients' needs, so trade-offs like added burden or potential limited utility of the data were mentioned. Future systems should be co-designed with patients and caregivers, as in recent human-centred research (e.g.,~\cite{Serban2025CTRAthletes}). We also highlight inequalities in prehabilitation access~\cite{Stewart2025InequalitiesPrehab} and obstacles in recruitment (28\% ineligible, 33\% prehabilitated outside hospital). Adding technology to this context can widen these inequalities (26\% thought the system would add burden to the program). Our findings reflect the views of those who accepted telemonitoring, were non-frail and enrolled in a hospital-based program. While they felt motivated by contributing to a life-saving effort, this sentiment may differ in routine care. Broader studies are needed to address diverse patient needs and real-world implementation.

\section{Conclusion}
This study underscores the importance of patient perspectives in designing telemonitoring systems for CRC prehabilitation. Participants appreciated the non-intrusive design and saw value in supporting their care team with useful data, which gave them a sense of agency. Yet, trade-offs emerged, including concerns about over-medicalisation and the need to respect autonomy. Social aspects, like involving caregivers, also proved valuable. While based on a limited group, these findings point to the need for broader research and co-designed systems that enhance prehabilitation while minimising burden and fostering trust.

\paragraph{Acknowledgements.} We are grateful for the valuable collaboration with the patients. We are also thankful for the support of our partners at Philips and Kinduct Technologies.

\appendix
\section{Appendix}
\subsection{Participants}
\label{participants}
\begin{table}[hbt!]\centering
\caption{Participant demographics}\label{tab:Experts}
\scriptsize
\begin{tabular}{p{1cm} p{1cm} p{1cm} p{2.3 cm} p{2.3cm} p{2.3cm}}\toprule
\textbf{ID} &\textbf{Sex} &\textbf{Age} &\textbf{Cancer stage} &\textbf{Living situation} &\textbf{Occupation} \\\midrule
P1 & M & 72  & 1 & With partner & Retired \vspace{0.5 cm} \\
P2 & M & 67  & 3 & With partner & Retired \vspace{0.5 cm} \\
P3 & M & 62  & 1 & With partner & Working (driver) \vspace{0.5 cm} \\
P4 & M & 80  & 1 & With partner & Retired \vspace{0.5 cm} \\
P5 & F & 67  & 1 &With partner & Retired \\
\bottomrule
\end{tabular}
\end{table}

\subsection{Interview Guide -- Example Questions} \label{interview}

\textbf{Experiences with the prehabilitation journey}

•	Things you enjoyed, things you found less enjoyable.

•	Challenges in managing physical activity and well-being according to prehab recommendations? Challenges when communicating with clinicians?

\textbf{Experiences with being remotely monitored}

•	Negative experiences, positive experiences, and motivators for engaging.

•	What did other people think about the sensors you wore? Did you wear it in social situations – did people make remarks? How did you feel afterwards?

\textbf{Using sensor data}

\textit{Show cards with sensor data types}

•	Interesting or irrelevant data? Why so?

\textit{Show cards with possible actions}

•	How could you use the data? At which moments in the prehab trajectory could you act on the data? How could the data enhance clinicians' feedback or your self-management? 

• In which way and for which purpose would you share data with your clinician? Would you be comfortable to do so?

•	Sharing data with family/partner– at which moments? For which purpose? 

•	Having an overview of the data – Weekly? Daily? Continue the measurement after surgery? 

•	With all this information available - from the sensors and from the diary - what kind of information or advice would you like to get more benefit from the program?

\textbf{Using self-reported data}

\textit{Show graphs of self-reported data}

•	Experiences of self-reporting, likes or dislikes. Did the diary make you think about things you would not normally think of? If you could use this data, how would it come in handy? 

\textbf{End of interview}

•	How could you see this telemonitoring system implemented in the prehab program? Further comments/points of view.

\subsection{Visual Materials} \label{visual}

\begin{figure}[H]
  \centering
  \includegraphics[width=0.7\linewidth]{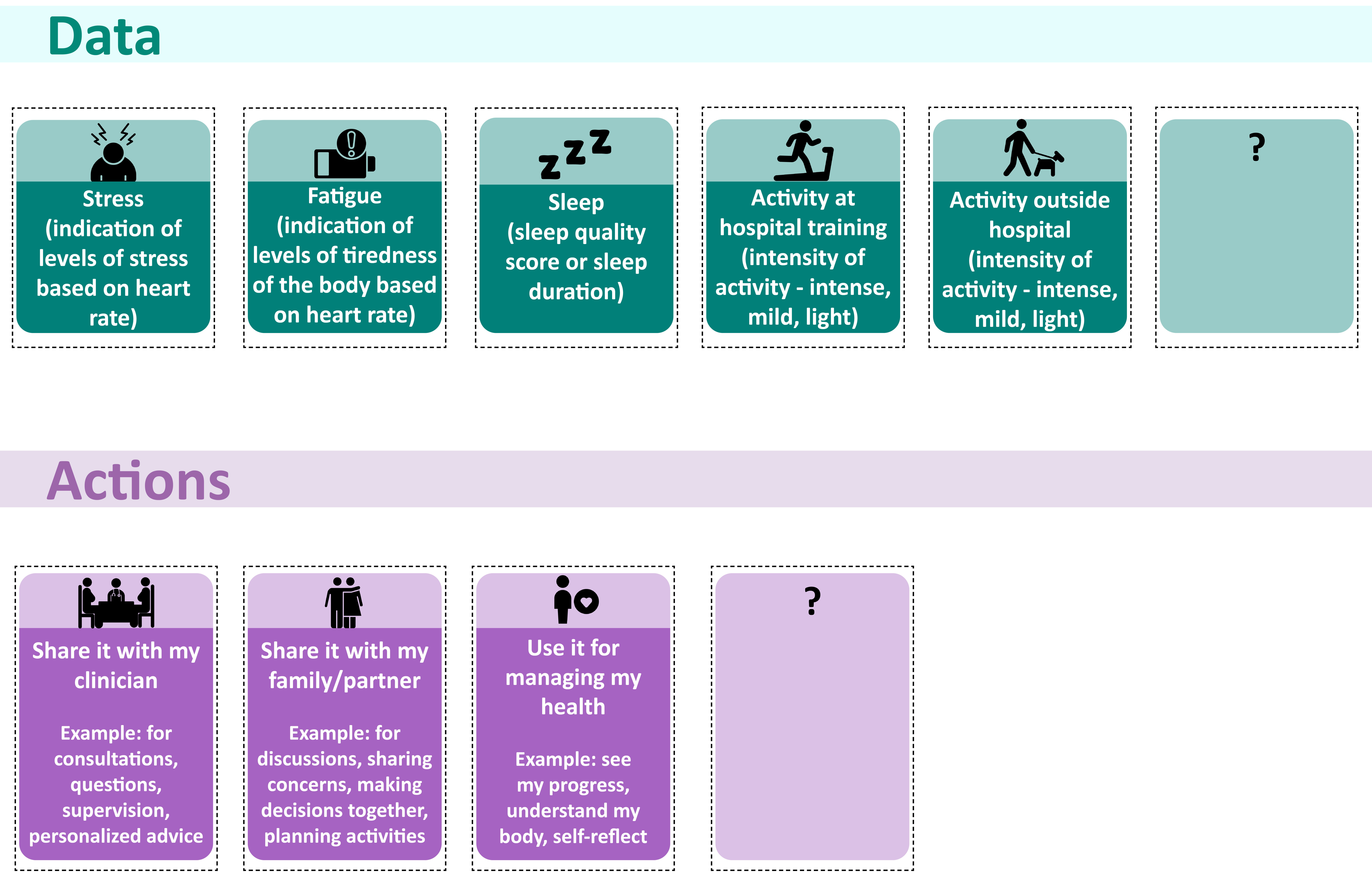}
  \caption{Visual aids in the form of cards representing potential data themes and actions. Empty cards left for participant suggestions.}
  \label{fig:infotype}
\end{figure}

\begin{figure}[H]
  \centering
  \includegraphics[width=\linewidth]{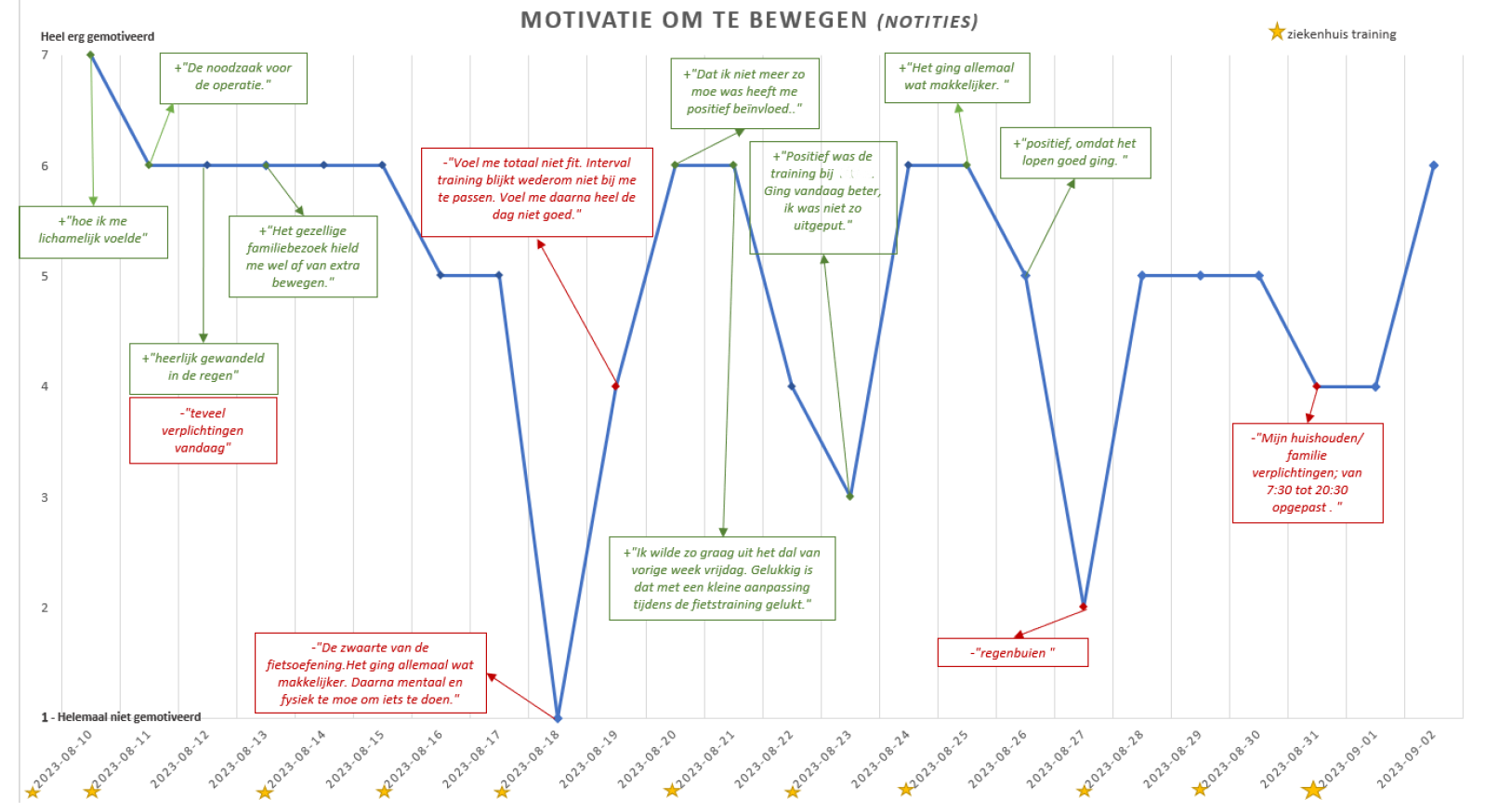}
  \caption{Example of a self-reported data visualisation showing P5's motivation for physical activity outside supervised training (marked with yellow stars), rated on a scale from 1 to 7, overlaid with perceived barriers (red boxes) and facilitators (green boxes) to physical activity.}
  \label{fig:infotype}
\end{figure}

%
%
%
\bibliographystyle{splncs04}
\bibliography{main}

\end{document}